# Application of photoreflectance to advanced multilayer structures for photovoltaics

D. Fuertes Marrón*, E. Cánovas**, I. Artacho

*Instituto de Energía Solar – ETSIT, Technical University of Madrid, UPM, Madrid, Spain.*

C.R. Stanley, M. Steer

*Department of Electronics & Electrical Engineering, University of Glasgow, UK.*

T. Kaizu, Y. Shoji, N. Ahsan, Y. Okada

*Research Center for Advanced Science & Technology, University of Tokyo, Japan.*

E. Barrigón, I. Rey-Stolle, C. Algora

A. Martí, A. Luque

*Instituto de Energía Solar – ETSIT, Technical University of Madrid, UPM, Madrid, Spain.*

* Corresponding author: dfuertes@ies-def.upm.es, tel: +34 914533573, Fax: +34 915446341

** Present address: Max-Planck Institute for Polymer Research, Mainz, Germany.

**Abstract**

Photoreflectance (PR) is a convenient characterization tool able to reveal optoelectronic properties of semiconductor materials and structures. It is a simple non-destructive and contactless technique which can be used in air at room temperature. We will present experimental results of the characterization carried out by means of PR on different types of advanced photovoltaic (PV) structures, including quantum-dot-based prototypes of intermediate band solar cells, quantum-well structures, highly mismatched alloys, and III-V-based multi-junction devices, thereby demonstrating the suitability of PR as a powerful diagnostic tool. Examples will be given to illustrate the value of this spectroscopic technique for PV including (i) the analysis of the PR spectra in search of critical points associated to absorption onsets; (ii) distinguishing signatures related to quantum confinement from those originating from delocalized band states; (iii) determining the intensity of the electric field related to built-in potentials at interfaces according to the Franz-Keldysh (FK) theory; and (v) determining the nature of different oscillatory PR signals among those ascribed to FK-oscillations, interferometric and photorefractive effects. The aim is to attract the interest of researchers in the field of PV to modulation spectroscopies, as they can be helpful in the analysis of their devices.

1.  **Introduction**

The complex electronic structure of novel photovoltaic devices demands characterisation tools able to test the quality of the material at an early stage of processing. If it were possible to choose those tools *à la carte*, the *ideal* characterisation method of choice would be fast, contactless and sensitive enough to probe the main optoelectronic properties of all types of nanostructures, multilayers and complex device designs, not requiring cryogenic cooling or vacuum/inert atmosphere during operation. Photoluminescence (PL) is one of the most widespread diagnostic techniques, providing useful information of radiative recombination processes controlled by kinetics of excited carriers, despite the fact that it normally requires cooling units assisted by high vacuum. Among the family of modulation spectroscopies, photoreflectance (PR) combines all advantages mentioned for the *ideal* characterization tool, appearing as robust complementary technique to PL and usually providing a wider range of information not easily accessible by other means. The aim of this work is to present recent results of the application of PR to a number of advanced photovoltaic devices of different technologies, including quantum-dot and quantum-well heterostructures, diluted nitrides, and multinary compounds. These examples serve to illustrate the main features of the technique and the samples under study as well as specific subtleties related to the interpretation of experimental results obtained.

PR belongs to the group of electro-modulation techniques, in turn a variety within modulation spectroscopies [1]. As such, it is basically an analogue method for taking the derivative of the optical spectrum of a material by modifying in some manner the measurement conditions by a perturbing agent [2]. The upper panel of Figure 1 represents the reflectance (R) spectrum of an exemplary semiconducting sample as a function of the photon energy. The step in R appears at the energy of a critical point in the electronic structure of the sample (associated with inter-band or inter-sub-band resonances; *i.e.*, a bandgap). The lower panels of Fig. 1 represent derivatives of first up to the third order of R with respect to the energy [3]. It can be observed that the derivative spectra somehow isolate the critical point out of R, neglecting minor changes of R away from the region of interest. The finite magnitude of R throughout the entire energy range avoids eventual indetermination of the derivative spectra, which could affect an analogue analysis based on transmittance of highly absorbing materials. Furthermore, it should be noticed that the sharpness of the signature in higher derivative orders of the spectrum is increased with respect to that of the first-derivative spectrum.

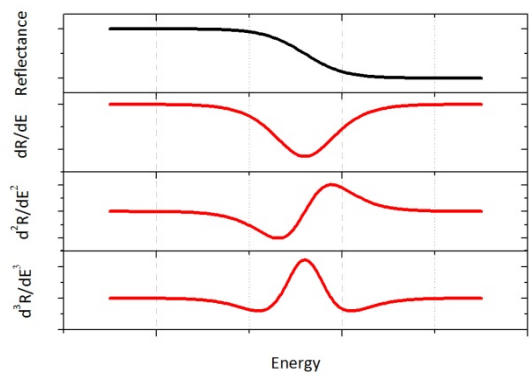

**Figure 1**: Reflectance spectrum (upper panel) and corresponding 1st, 2nd and 3rd derivative spectra with respect to the energy (lower panels) of a typical semiconducting sample.

Electro-modulation uses an electric field as a perturbing agent. In particular, PR relies on the diffusion of optically generated charge carriers (without the need of electrical contacts) to modulate electric fields and change the band bending associated with space-charge regions, typically present at surfaces and interfaces. The dielectric constants of the sample, and thereof its reflectance and transmittance, are thus perturbed upon application of a periodic modulation on the electric field. Photoreflectance measures tiny changes in reflectance upon the action of the perturbing agent and is typically referred to as ∆R/R ratios. As such, it follows that PR signatures are closely related to the derivative of the dielectric constant of the material under study, which in turn governs its reflectance. This point will be discussed in more detail in the following sections, where a brief description of the theoretical basis of PR and modulation spectroscopy is reviewed. It is worth mentioning that currently available phase-sensitive detection systems can obtain PR-signals as low in magnitude as $10^{-6}$-$10^{-7}$.

## 2. Materials and methods

The PR-setup used in this work is shown in Figure 2. The light beam of a quartz-tungsten-halogen (QTH) lamp (up to 250 W) is passed through a monochromator (1/8 m Cornerstone-Newport) and focused by optical lenses on the surface of the sample with an intensity $I_0(\lambda)$. The light directly reflected $I_0(\lambda)R(\lambda)$ is focused with lenses and measured with a solid-state detector (Si or Ge). The current signal is pre-amplified (Keithley) and transformed to a voltage (dc-signal). Superimposed onto the light spot at the surface of the sample, a laser beam chopped mechanically at a certain frequency (in most of our measurements 777 Hz) provides the modulated perturbation. Laser lines at 325 nm and 442 nm of a 10 mW He-Cd laser and at 632.8 nm line of a 17 mW He-Ne laser have been used in this work. The energy of the pump laser beam determines the highest critical point accessible of the sample. The signal recorded at the detector contains two components: the dc average signal $I_0(\lambda)R(\lambda)$ and the ac modulated signal $I_0(\lambda)\Delta R(\lambda)$, where $\Delta R(\lambda)$ is the change in reflectance caused by the modulated perturbation. The signal feeds a lock-in amplifier (Stanford Instruments), which tracks the ac signal at the given frequency. The relative change in reflectance is then obtained by normalizing the ac signal with respect to the dc component in a computer, giving typical values between $10^{-3}$ and $10^{-6}$. Such normalization represents an additional advantage of the technique in terms of eventual incomplete light collection: provided enough light reaches the detector, the line-shape of the modulated spectrum is not affected by the loss of some light in $I_0$.

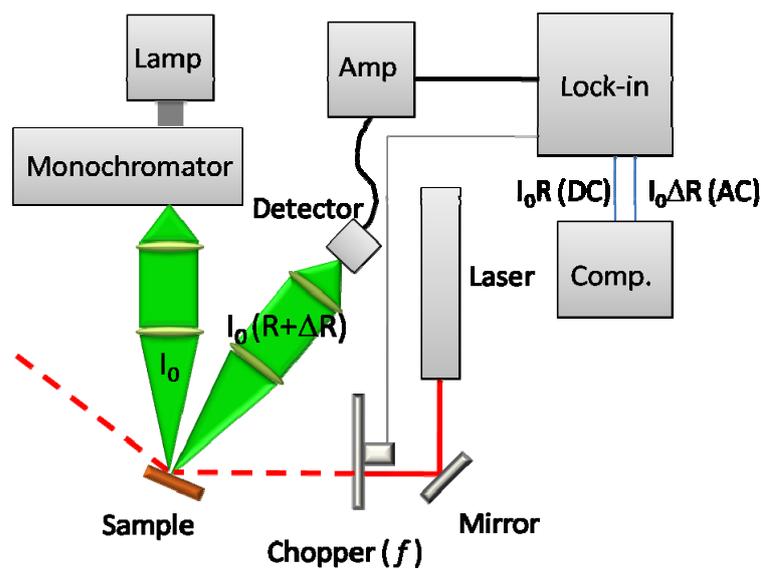

**Figure 2**: Experimental setup for PR measurments.

The quality of the PR signal obtained in a measurement carried out as described in the Fig. 2 is typically affected by two spurious signals. First, the possibility that scattered laser light reaches the detector cannot be excluded. Ideally, light collected at the detector should be that of the reflected probe beam originated at the lamp, with no contribution from the laser pump beam. In practice, however, it is worthwhile to include pass filters in front of the photodetector as to avoid any residual scattered laser light contributing to the signal, precisely at the frequency of the modulation. Second, the eventual luminescence of the sample upon super-bandgap laser excitation may contribute a spurious ac-signal, again at the precise frequency of the modulation. The impact of luminescence can be reduced by increasing focal distances of the collecting optics or else by incorporating a monochromator before the detection unit [4]. More sophisticated approaches include the chopping of the probe beam at a different frequency as the pump beam, setting then the detection at the locking to track the sum (or difference) of both frequencies [5]. If not properly attenuated, both laser scattering and sample luminescence will contribute apparent white noise to the PR spectra. If sufficiently intense, they can eventually bury the actual signal in a high level background and therefore care should be taken to avoid their contributions.

3. Theory

The linear optical response of a medium to electromagnetic fields is governed by its dielectric constant, which describes the response of electrons to the field. Considering rather general conditions (non-magnetic media, transverse electromagnetic radiation, frequency of radiation much larger than interatomic distances, and causality) [1], Kramers-Kronig relations apply linking the real and imaginary parts of the dielectric function (or alternatively of its refractive index) and the problem of electrons in a solid responding to the field can be solved leading to a general expression of the dielectric constant of the type $\varepsilon_{jj}$ = 1 + {interband contributions} + {intraband contributions} [6]. Interband contributions may include both excitonic and non-excitonic transitions, leading to two different situations in the line-shape analysis of PR, as described later, and are dominant in PR spectra. Knowing the dielectric constant or the refractive index, the absorption coefficient, reflectivity and transmittivity coefficients can be derived from them.

Modulation spectroscopies rely on the ability of perturbing the dielectric constant $\varepsilon = \varepsilon_1 + i\varepsilon_2$ of the medium by some means, in an amount $\Delta\varepsilon = \Delta\varepsilon_1 + \Delta\varepsilon_2$. From the general relation between reflectance and dielectric constant:

$$R = \left|\frac{n-n_a}{n+n_a}\right|^2,$$

$$n^2 = \varepsilon,$$

where subindex $a$ refers to ambient considering the example of a front surface exposed to air, and $n$ is the refractive index, we can express relative changes in reflectance as [7]:

$$\frac{\Delta R}{R} = \alpha \Delta\varepsilon_1 + \beta \Delta\varepsilon_2.$$

Similar considerations apply to internal interfaces. $\alpha$ and $\beta$ are referred to as the Seraphin coefficients [8]. Seraphin coefficients have a characteristic variation in semiconductors: below the gap, the imaginary part of the dielectric function can be neglected; above the bandgap, ΔR/R always includes contributions from real and imaginary terms. Additionally, Seraphin coefficients are known to be sensitive to optical interference effects and consequently one should be cautious when neglecting imaginary terms in multilayer structures like quantum-well superlattices [9]. Functional forms for $\Delta\varepsilon_1$ and $\Delta\varepsilon_2$ can be calculated provided that the dielectric function and the type of critical point are known. The general expression for $\varepsilon$ in the one-electron approximation considering Bloch states with energy $E(k)$, where $k$ is the wave vector, can be expressed as [7]:

$$\varepsilon = Ce^{i\theta}(E - E_g + i\Gamma)^n,$$

assuming the parabolic band approximation at critical points, where $C$ is the amplitude, $\theta$ a phase projection factor, $E_g$ the absorption threshold, $\Gamma$ a phenomenological broadening parameter related to Lorentzian lifetime broadening due to carrier scattering, and $n$ is an exponent related to the type of critical point involved. The type of critical point can be deduced from the parabolic band approximation:

$$E(\vec{k}) = E_g + \frac{\hbar^2 k_x^2}{2m_x} + \frac{\hbar^2 k_y^2}{2m_y} + \frac{\hbar^2 k_z^2}{2m_z} + \ldots,$$

from which it follows that 3D-critical points contain three finite quadratic terms in the expansion, and 2D- and 1D-critical points refer to situations when either one or two effective masses along principal axes (labeled x, y, and z), respectively, are very large. According to the functional form of $\varepsilon$, it is found that n=-0.5 for 1D-, n=0 for 2D-, and n=0.5 for 3D-critical points; additionally, n=1 for excitonic and confined states, for which the parabolic band approximation breaks up and states are essentially non-dispersive [7].

Electroreflectance is a particular case of modulation spectroscopy, as compared to piezo- or thermo-reflectance, since it relies on the perturbation of the dielectric constant by an electric field. Such an external field can have two main effects in a piece of semiconductor. It can induce band tilting and therefore carrier acceleration, as opposed to bandgap modulation by strain or temperature cycles. This situation is expected in semi-insulating samples and intrinsic material. Most likely, the external field can act upon already existing built-in potentials, either at free surfaces or at interfaces between different layers making up the sample, modulating the band bending and consequently the overall field present in the sample. This is typically the case in PR, where the modulation of the field is achieved by photo-generation of carriers that are separated at space-charge regions. The periodic photovoltage induced by the perturbing light source changes the bending of the bands and modulates built-in potentials. In any case, electron-hole pairs in a field experience a potential which depends upon position, translational symmetry, that otherwise is preserved in quasi-neutral regions of the semiconductor, is thus destroyed at space-charge regions and mixing of wave functions of different k-vectors occurs there. It was shown by Aspnes [10-12] that such type of band-tilting modulation in bulky semiconductors is directly related to the third derivative of the unperturbed dielectric constant with respect to the energy (a simpler, qualitative description of this result is given in Refs. [13-14]), whereas bandgap modulations can be classified as first-derivative analogues.

Consequently, the line-shapes observed in photo-reflectance can be directly obtained from the functional form of $\varepsilon$. Qualitative differences in line shapes of first- and third-derivative nature can indeed be inferred from Fig. 1. The third-derivative regime is actually achieved only if the energy gained by the electron upon acceleration in the presence of the field is less than the broadening parameter described above. This is the so-called *low-field* regime, where broadening effects dominate. The characteristic energy related to the presence of the field which determines whether or not the low-field regime is operative is referred to as the electro-optic energy $\hbar\Omega$:

$$\hbar\Omega = \left(\frac{q^2\hbar^2F^2}{8\mu}\right)^{1/3},$$

where $q$ is the elementary charge, $F$ is the field intensity, $\mu$ is the joint density of states effective mass in the direction of the applied field. When the field strength is higher, such that acceleration dominates now over broadening effects ($\hbar\Omega \gtrsim \Gamma$), the third-derivative character of the modulated reflectance does no longer apply as we enter the *medium-field* regime, dominated by Franz-Keldysh effects that will be described later in Section 4.5. Finally, the *high-field* regime refers to the situation in which the field is so intense that significant band tilting occurs over distances comparable to the unit cell, altering the band structure itself (Stark effect).

From the functional form of $\varepsilon$ it can be shown that third-derivative spectra in the low-field regime exhibit line shapes equivalent to the functional form of $\varepsilon$ with a new exponent factor $m$ defined as $m=n-3$. Therefore, the allowed values of $m$ are now $m=-2.5$ for a 3D-, $m=-3.0$ for 2D-, $m=-3.5$ for 1D-critical points, and $m=-2$ for excitonic transitions. The resulting functional is referred to as the third-derivative functional form (TDFF) [12] and has become the fundamental tool for analysis of PR line shapes. An application of TDFF-based analysis will be shown in the next section.

In relation to the previous comment referring to Fig. 1, it should be mentioned that the identification of PR signatures as first- or third-derivative line shapes is usually not straightforward at first sight. To illustrate this point, Figure 2 shows an example of the same TDFF for three different values of the phase factor (in radians), keeping the critical point energy, the broadening parameter, the amplitude, and the *m*-exponent constant. Fitting procedures based on minimum squares are normally required in the analysis of PR line shapes in order to ascribe the nature of the transitions to the observed signatures.

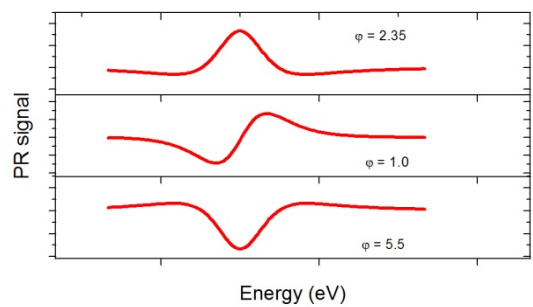

**Figure 3**: Simulated third-derivative line-shapes of a PR signal for different values of the phase factor, with all other parameters of the TDFF fixed.

With PR we can therefore obtain detailed information about the existence of bandgaps and their nature ($E_g$ and *m*-exponent of TDFF); the Lorentzian broadening of the line-shapes ($\Gamma$); the bound or unbound character of the electronic states involved in transitions (from its derivative order and *m*-exponent); and the electro-optic energy and the magnitude of associated electric fields at surfaces and interfaces, among other properties. PR can be performed at room temperature, on wafer-sized material or locally at sub-millimeter scale, even *in-situ* during sample processing. It is non-destructive and typically does not require vacuum or an inert atmosphere. Furthermore, the diffusion of optically generated carriers allows buried interfaces far beyond the penetration depth of the optical pumping to be probed.

## 4. Results

### 4.1 Quantum dot-based solar cell prototypes

Quantum-dot based solar cells have been proposed as candidates for the realization of intermediate-band solar cells (IBSCs) [15]. The basic idea is illustrated in Fig. 4, where the confined states of a quantum dot allow two additional absorption processes $E_1$ and $E_2$ to occur at energies below the nominal bandgap of the host semiconductor, in conjunction with band-to-band $E_0$ transition. The expected PR response of the schematic structure on the left as a function of photon energy is shown on the right panel. As a result of additional absorption onsets, higher photocurrents over those generated at the host material alone are expected. If voltage preservation can be ensured by an adequate device design which includes conventional semiconductors acting as emitters on either side of the QD-containing material [15], the resulting solar cell is expected to yield efficiency figures well above the Shockley-Queisser limit of conventional PV devices [16].

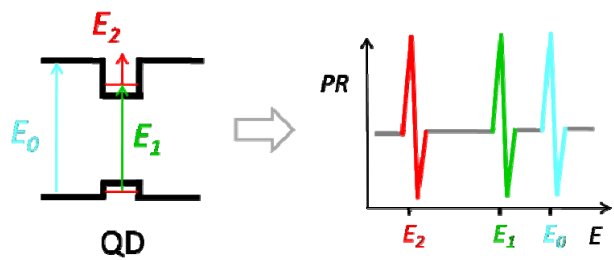

**Figure 4**: (Left) Optical absorption thresholds $E_1$ and $E_2$ in a QD embedded in a semiconductor host, where band-to-band transitions are identified as $E_0$. (Right) Schematic PR spectrum as a function of photon energy expected from the structure on the left.

One important issue in the optimization of such a device is the appropriate tuning of the position of the intermediate band, stemming from the confined ground state of the nanostructures, with respect to the band edges of the host material. First prototypes of QD-IBSCs were fabricated by embedding self-assembled InAs QDs in GaAs as active material [17] and PR was used in the analysis of the structures [18]. The resulting band alignment, however, is not optimal for IBSC operation and a large interest is driving research towards enlarging the bandgap of the host material and the energy difference between confined states at the QDs and the conduction band edge of the host. A way to go is by replacing GaAs by AlGaAs as host material, taking advantage of the possibility of tuning the band gap of the AlGaAs host as a function of Al-content. The PR spectrum of the structure schematically shown in Figure 5 (left), obtained under 325 nm laser line pump chopped at 777 Hz and using a Si-detector, is shown in the same figure (center). The structure, grown by MBE, contains 20 layers of InAs QDs embedded in $Al_{0.25}Ga_{0.75}As$, with spacers including 2 nm of $In_{0.2}Al_{0.2}Ga_{0.6}As$ capping plus 6 nm and 44 nm $Al_{0.25}Ga_{0.75}As$ grown at low and high temperature (LT, HT), respectively. The structure ends up with a QD layer at the front surface as designed for AFM studies.

The fundamental gap of $Al_{0.25}Ga_{0.75}As$ is clearly resolved by PR at 1.74 eV, with an additional signature assigned to spin-orbit splitting at 2.1 eV. The solid line in Fig. 5 shows the fit according to the TDFF described before, including up to 8 critical points within the energy range 1.25 – 1.82 eV. Resulting fitting parameters are collected in Table 1 and the corresponding residuals are included in the inset of Fig. 5 (center). We notice the extended-state character of signatures from the $Al_{0.25}Ga_{0.75}As$ host (*m*-exponent of 2.5) in contrast to sub-bandgap signatures (non-3D critical points). The dashed area in the figure highlights the range of sub-bandgap signatures, attributed to confined states from the dots and the corresponding wetting layers. The right panel in Fig. 5 compares the low energy range of the PR measurement carried out with a Ge-detector and the corresponding PL spectrum obtained at room temperature from the same sample. Up to five contributions could be resolved by comparing PL and PR spectra (dotted lines on right panel) which likely correspond to ground and lowest excited states of the dots. The signal-to-noise ratio of PR recorded in this energy range did unfortunately not allow any reliable fitting. Additional signatures are readily visible around 1.3 eV in the PR spectrum which are PL-silent, so is the rest of sub-bandgap signatures in the dashed area of the measurement performed with the Si-detector (right figure) and the signatures related to the $Al_{0.25}Ga_{0.75}As$ host.

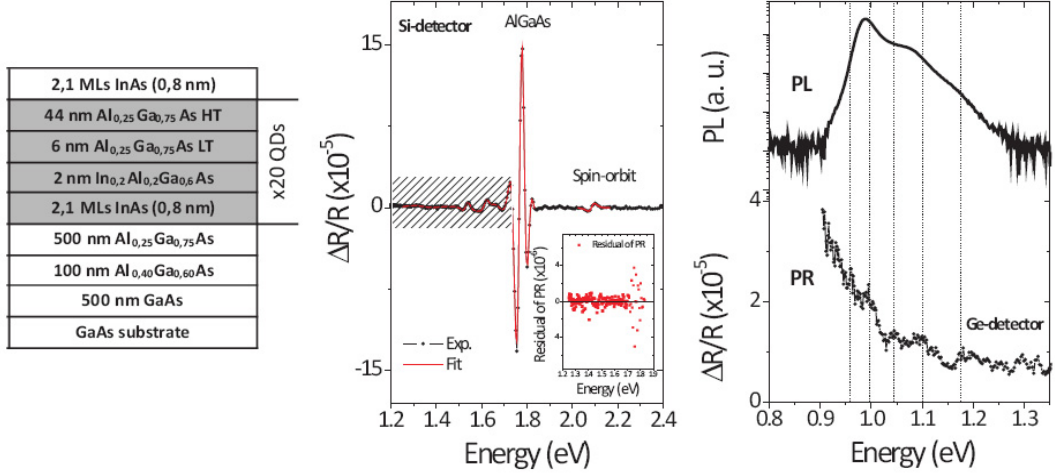

**Figure 5**: (Left) Scheme of a MBE-grown sample containing 20 layers of InAs QDs embedded in $Al_{0.25}Ga_{0.75}As$, with spacers including 2 nm of $In_{0.2}Al_{0.2}Ga_{0.6}As$ capping plus 6 nm and 44 nm $Al_{0.25}Ga_{0.75}As$ grown at low and high temperature (LT, HT), respectively. The structure ends up with a QD layer at the front surface, as designed for AFM studies. (Center) PR spectrum (dots: experimental; solid: TDFF fit) of the structure on the left, recorded with 325 nm laser line pump chopped at 777 Hz and 150 W QTH lamp, using a Si-detector. The inset shows the residuals of the fit. (Right) Comparison of room-temperature photoluminescence (upper panel) and PR (lower panel, recorded with a Ge-detector) over the energy range 0.80-1.35 eV. Dotted lines correspond to signatures ascribed to confined states at the dots. The apparent increasing background at lower energies is due to the diminishing intensity of the probe beam at the corresponding energies and the different noise levels recorded between AC- and DC-signals, causing the ΔR/R ratio to diverge.

**Table 1**: Fitting parameters of the PR spectrum shown in Fig.5 according to the TDFF. Critical point energy (E), broadening factor (Γ) and *m*-exponent are included for nine PR signatures, including sub-bandgap and host-related features.

|  | Sub-bandgap signatures (>1.33 eV) | | | | | AlGaAs | | | S-O |
| --- | --- | --- | --- | --- | --- | --- | --- | --- | --- |
| **E (eV)** | 1.34 | 1.52 | 1.53 | 1.62 | 1.72 | 1.74 | 1.77 | 1.80 | 2.09 |
| **Γ (meV)** | 59 | 36 | 60 | 30 | 83 | 38 | 30 | 25 | 28 |
| **m** | 1.8 | 2.0 | 3.2 | 3.7 | 2.1 | 2.5 | 2.5 | 2.5 | 2.5 |

These results indicate that: (i) it is possible to enlarge the energy difference between the ground state of the confined structures and the absorption edge of the host material to about 0.8 eV (in fact lowered by about 100 meV by the presence of the wetting layers); and (ii), that this can be achieved but likely at the expense of introducing a relatively large density of discrete density of states associated to highly excited states of the QDs. Further optimization is therefore required in order to fully exploit the favorable band diagram expected from this structure.

*4.2 Multi-quantum well structures: Probing excited states*

Glembocki et al. have first showed that PR could be used to obtain derivative-like spectra from multiple quantum wells (MQW) [19]. Since then, the technique has been widely applied to a number of nanostructures (see, for example, Ref. [20]). Figure 6 shows a comparison of PR and PL spectra taken at room temperature of a quantum well structure consisting of four different $In_{0.32}Ga_{0.68}As$ wells of 11 nm, 8.5 nm, 6.5 nm, and 4.5 nm width, respectively, separated by GaAs spacers, all embedded in an $Al_{0.40}Ga_{0.60}As$-GaAs superlattice host replicating a $Al_{0.20}Ga_{0.60}As$ matrix, as schematically shown in the left panel.

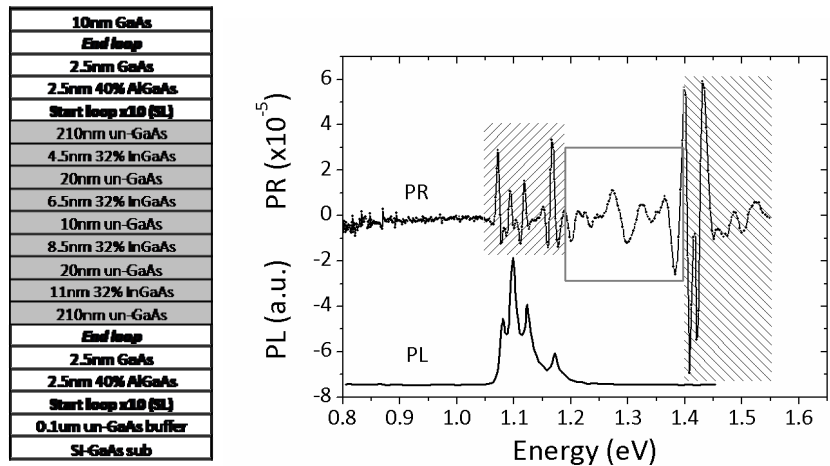

**Figure 6**: (Left) Scheme of a MBE-grown sample containing four $In_{0.32}Ga_{0.68}As$ quantum wells of 11.0, 8.5, 6.5 and 4.5 nm width separated by GaAs spacers and embedded in $Al_{0.40}Ga_{0.60}As$/GaAs superlattice. (Right) PR spectrum obtained from the sample on the left (632.8 nm laser line chopped at 777 Hz, 150 W QTH lamp, Ge-detector).

PL shows four prominent signatures between 1.08 eV and 1.17 eV, attributed to the four ground states of the four wells. The rest of the sample is apparently PL-silent. PR shows a much richer structure which includes not only the luminescent fundamental states (dashed area on the left) but also the capping and host material (dashed area on the right) and sub-bandgap signatures related to excited states of the wells (rectangle). The location of confined signatures can be correlated to theoretical calculations of the model quantum mechanical potential wells when material properties and dimensions are known, and it should be possible to assign each of the recorded signatures with their corresponding calculated transitions. Nevertheless, at the sight of Fig.6, it becomes obvious that PR can facilitate reliable information about sub-bandgap transitions that do not show up in PL spectra under the same measuring conditions. The importance of this result concerning PV-applications lies currently on the interest of utilizing QWs as a means to finely tune the current matching between sub-cells in multijunction configurations [21], where different solar cells with different absorption thresholds are stacked in a series connection. Any rough estimation of the current gain expected by the incorporation of QWs in a photovoltaic device based on PL, like the spectrum shown in Fig. 6, will severely underestimate the actual photocurrent stemming from the confined structures, as it will not consider contributions from excited states. PR offers a probing technique sensitive to absorption rather than recombination processes and therefore much better suited for photocurrent generation analysis.

*4.3 Dilute nitride: Probing the band-anticrossing model*

A means of increasing the number of stacking layers during the growth of QD-containing structures is to introduce strain-compensating layers as spacers between the QD-planes. The compensation of strain inhibits the appearance of threading dislocations interrupting the self assembly of subsequent QD-stacks and helps maintaining good electronic properties of the device. In the system InAs/GaAs, phosphorous is normally supplied during the growth of epitaxial layers, acting as strain relief when incorporated into the spacer between QD-layers. An alternative approach has been suggested using diluted nitrogen in GaAs in compositions close to $Ga_{0.99}N_{0.01}As$ ([N]=1 %at) that has allowed the growth of dislocation-free structures of more than 100 layer stacks [22]. Figure 7 (left) shows the stacking layer sequence of a MBE-grown sample containing 50 layers of InAs QDs with $Ga_{0.99}N_{0.01}As$ spacers. The nearly ideal lattice matching of the diluted nitride with the GaAs host in this sample is demonstrated in the XRD diffractogram of Figure 7 (right), showing the rocking curve around the (004) diffraction peak of GaAs and its shoulder at low ω due to the diluted nitride. The superlattice arrangement of the spacer layers is reflected in the subsidiary peaks evenly distributed at higher and lower ω–angles.

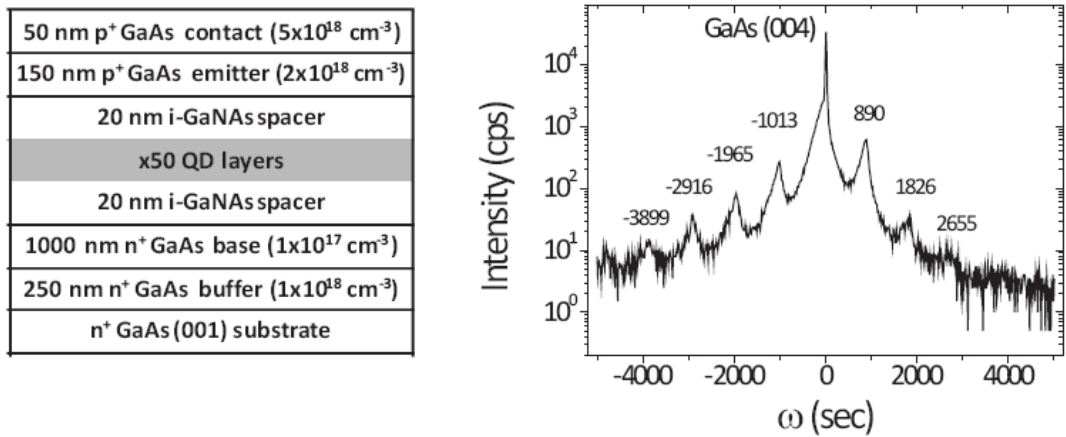

**Figure 7**: (Right) Schematic representation of the growth sequence employed for a 50x InAs-QD layer stack capped with $Ga_{0.99}N_{0.01}As$ spacers in a GaAs host. (Right) XRD-rocking curve of the epitaxial structure around GaAs (004) reflection confirming the quality of the structure and the nearly optimal lattice match between GaAs and the diluted nitride. Susbsidiary peaks reflect the superlattice arrangement of the multilayer stack.

GaNAs belongs to the so-called highly mismatched alloys, characterized by the presence of a highly electronegative and light element in a semiconductor matrix [23]. The presence of the highly electronegative species dramatically alters the electronic structure of the host material, resulting in the splitting of the band edge of the host into two branches as described by the band-anticrossing model (BAM) [24]. The origin of the splitting following the BAM model is explained by a similar symmetry character of electronic levels introduced by the impurity which are nearly resonant with one of the band edges of the host. As a result of the interaction between electronic states from the host and the impurity, the system arranges the states in a similar fashion as described by the theory of molecular orbitals (TMOs) upon formation of molecules from isolated atoms. Within this theoretical frame, a new bunch of electronic states forms the $E_-$ band (bonding states in terms of TMOs) and the rest of electronic levels is rearranged into the $E_+$ band (analogue to the antibonding branch in TMOs). This phenomenon is illustrated in Figure 8 [25].

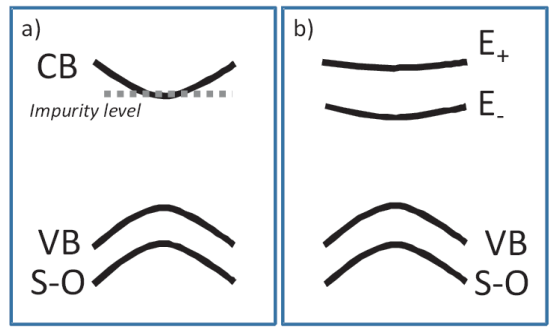

**Figure 8**: Schematic representation of the band-anticrossing model [23]. (Left) Dispersion relation E-k before the interaction of VB- and CB-states of the semiconductor host with impurity levels nearly resonant with CB-states. (Right) After interaction, symmetry properties of nearly resonant states force the splitting of original states into two branches $E_+$ and $E_-$ of allowed states, opening a bandgap between them.

Figure 9 shows PR spectra (325 nm line of HeCd laser, 90 W QHL) of the 50xQD layer stack sample presented in Figure 7. The PR overview (Fig. 9 center) shows prominent signatures of GaAs, identified as the fundamental gap $E_0$, the spin-orbit transition $E_{SO}$ and the high energy $E_1$ critical point. Measurements were performed under polarized light according to the scheme shown in Fig. 9 (right). Differences in intensity in the PR spectra are clearly seen for the three main contributions assigned to GaAs. Polarization dependence in the response attributed to $E_1$ has been used to determine the optical anisotropy related to (001) GaAs-terminations originated from the arrangement of top-most As-Ga bonds (along [110] in Ga-terminated or along [1-10] in As- terminated surfaces) [26]. No further differences in intensity of PR signals related to the polarization character of the probe beam were expected at $E_0$ or $E_{SO}$ though. However, differences in intensity between *s*- and *p*-polarized spectra are evident, as can be seen in the enlarged region of the left figure. Such a polarization dependency is attributed to the presence of the diluted nitride in a multilayer stack, acting as spacer in the QD-structure, a dependency which is particularly notorious in the signatures recorded between 1.2 and 1.3 eV. In addition, the overview spectrum on the right panel shows PR-signatures attributed to the electronic structure of the nitride that have been identified according to the BAM as $E_-$ (1.2 eV), $E_-+\Delta_{SO}$ (1.54 eV), and $E_+$ (2.0 eV) in good agreement with previous results [27], although we cannot completely rule out subsidiary Franz-Keldysh oscillations (FKOs) affecting to some extent the apparent position of signatures above $E_0$(GaAs). It should be mentioned that PR has significantly contributed to the development of the BAM, as described in the work of Walukiewicz and co-workers [23-25] and is one of the techniques routinely used for testing highly mismatched alloys. Notice that the $E_-$ signature in Fig. 9 (left) shows a double feature separated by some 65 meV, that has been tentatively attributed to the splitting of the heavy and light holes due to the compensating strain undergone by the GaNAs layers [28]. The spectrum also shows the signature of what is believed the ground state of the QDs at 1.02 eV.

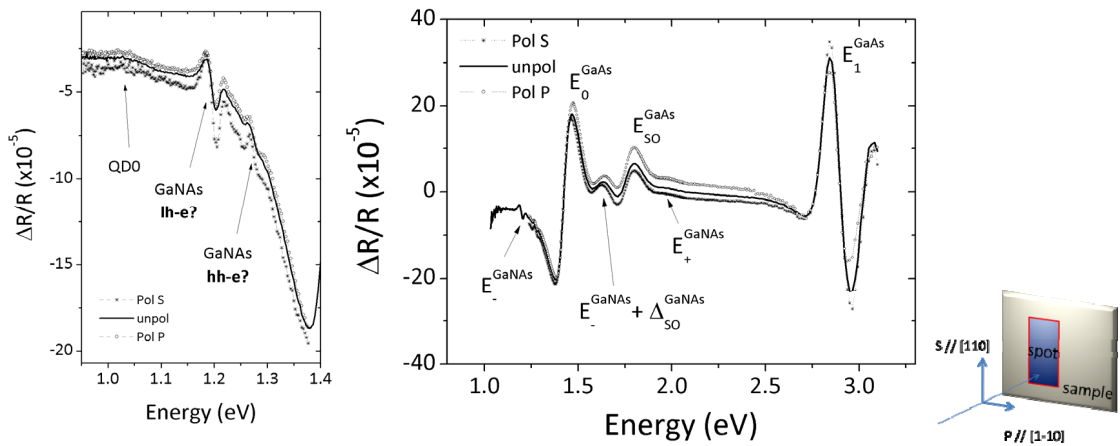

**Figure 9**: PR spectra of the 50xQD layer stack with diluted nitride spacers of Figure 7. (Left) Enlarged view of the low energy range (Ge-detector), where signatures form QDs and GaNAs are clearly resolved. Signatures around 1.20 and 1.26 eV are attributed to light and heavy hole contributions of the $E_-$ branch. The prominent signal close to 1.4 corresponds to the fundamental gap of the GaAs host. (Center) Overview PR spectrum (Si-detector) of the same sample showing identified signatures attributed to the GaAs host and the GaNAs spacers. For the latter, assignment has been performed following the band-anticrossing model. (Right) Polarization scheme used in the PR spectra, referring *s*- and *p*-polarization to the crystallographic orientation of the sample. Open symbols in the spectra correspond to *p*-polarization, solid symbols to *s*-polarization.

### *4.4 Oscillations below the bandgap: Simple interference or photorefractive effect?*

In addition to the results of the previous section, the same QD-structure with diluted nitride as spacer illustrates a remarkable effect that has been frequently found to complicate the analysis of PR under specific conditions. Figure 10 shows PR spectra of the same sample as before in the energy range between 0.8-1.4 eV, recorded with a Ge-detector using two different laser sources (632.8 nm and 325 nm lines of a He-Ne and a HeCd laser, respectively) and two different lamp intensities (90 and 230W). The first thing to notice is the independence of the PR spectrum of the intensity of the lamp used, when comparing spectra at 90 and 230 W. However, changing the laser excitation from 325 nm (as in Fig. 9) to 632.8 nm activates conspicuous sub-bandgap oscillations below 1.2 eV. The origin of the oscillatory behavior of PR signals below absorption thresholds is normally related to Fabry-Perot interference effects due to multiple reflections at the front and rear surfaces of thin films if the thickness of the layers is of the same order as the wavelength of the reflected light. In the structure under study, each GaNAs layer is 20 nm thick with a total of 51 layers (the QDs intercalated in-between), resulting in a total thickness slightly above 1µm. The layer thickness of the diluted nitride is therefore within the range of the wavelengths at which the interference effects show up in the spectra, below the absorption threshold set by the lowest $E_-$ branch around 1.2 eV. The interesting issue in this case is that no evidence of Fabry-Perot interference appears in the corresponding DC spectra (see Fig. 10 right), ruling out the possibility of a purely geometrical effect. Interference effects resulting in oscillatory signals with identical periods both in AC and DC signals (although with different phases) are routinely observed in thin transparent layers and are attributed to multiple reflection of the probe beam. The interference effect shown in Fig. 10 is related solely to the periodic perturbation at the ac-frequency and results from the laser perturbation of the sample. Such photo-refractive effect has been demonstrated before [29-34] and results from the change in dielectric constant induced by the laser on the sample under study. Such variation of the refractive index can in turn facilitate the fulfillment of the interference condition. It is thus evident that one should be cautious when interpreting PR results in transparency ranges of hosts showing sub-bandgap signatures when nanostructures are present in the samples. It could be tempting to ascribe sub-bandgap signatures to embedded nanostructures, but the possibility of photorefractive effects governing the ac-response should be excluded first. Furthermore, spurious oscillatory effects could mask the presence of actual signatures related to the embedded nanostructures. In such cases, a Fourier analysis of the periodic signal is recommended in order to resolve buried tiny structures. Cross-checking experiments using two different excitation wavelengths as pump beams, as shown in Fig. 10, are appropriate in order to distinguish eventual oscillatory effects. It should be noticed, however, that the absorption depth of the pump beam plays an important role in the magnitude of the observed PR-signatures, as can also be seen in Fig. 10. In this case, the use of 632.8 nm laser line as pump with a larger absorption length and thus penetrating deeper into the device structure enhances the response of the diluted nitride, as inferred by a higher intensity at related critical points, whereas signatures ascribed to GaAs show a reduced intensity.

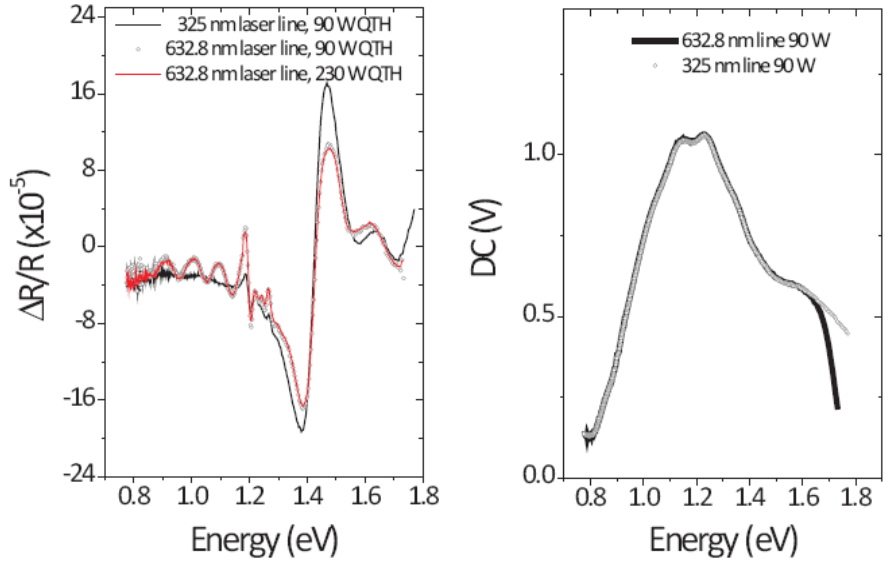

**Figure 10**: (Left) PR spectra of the 50xQD layer stack with diluted nitride spacers of Figure 7, using unpolarized light with 325 nm and 632.8 nm pumping laser lines, 90 and 230 W QTH lamp intensity and Ge-detector. (Right) Corresponding dc-siganls recorded simultaneously, showing no evidence of oscilations below 1.2 eV.

*4.5 Oscillations above the bandgap: Franz-Keldysh effect and the medium field regime.*

An electric field present in a semiconductor can modify its light absorption properties relative to the situation without field. The effect of the field is schematically shown in Figure 11 and is referred to as the Franz-Keldysh effect [35,36]. In equilibrium with no field present, the electronic states associated to valence and conduction bands can be described by Bloch functions as delocalized states with the translational properties of the crystal lattice embedded into it. When an electric field is applied to the semiconductor, the translational symmetry of the crystal is broken, resulting in band tilting in the corresponding band diagram. As a result, charge carriers are accelerated by the field. Three additional effects occur in the near band edge region, namely (i) the electronic states are described by Airy functions, rather than by Bloch functions, accounting for the return point at the band edges in the corresponding quantum-mechanical finite potential well problem; (ii) a de-phasing of the wave functions occur, which to a first order is proportional in some manner to the band tilting and thus to the intensity of the electric field (see levels E1, E2 in the Fig. 11); and (iii) due to the finite potential well felt by the carriers, the wave functions do not vanish at the classical return points but instead they leak into the bandgap with a typical exponential decay. Resulting from (ii) and (iii), two observable consequences are derived. First, the absorption threshold of the semiconductor under an electric field is shifted to lower energies, as the band tilting resulting from the field now allows photo-assisted tunneling via the leaking tails of the wave functions within the nominal bandgap. And second, the de-phasing in wave functions results in an oscillatory behavior of optical absorption at energies above the nominal bandgap of the semiconductor, the so-called Franz-Keldysh oscillations (FKOs), the period of which is proportional to the 2/3 power of intensity of the electric field, as discussed a bit later. FKOs are rapidly damped, showing a characteristic decay for increasing energy above the absorption threshold.

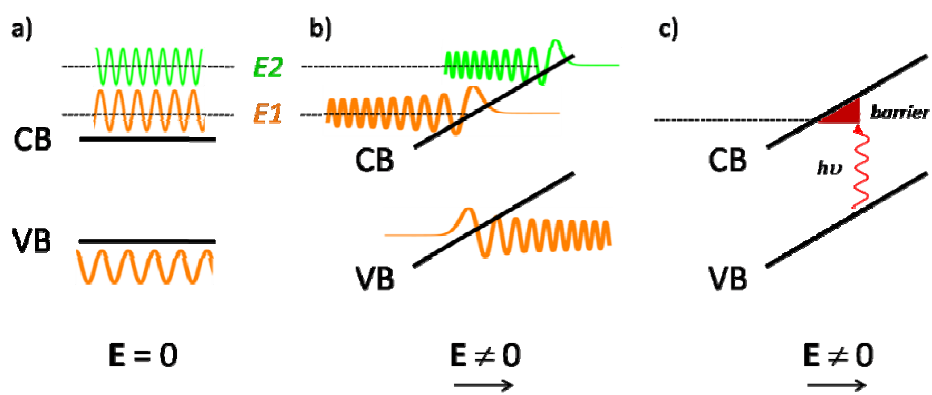

**Figure 11**: (Left) Band diagram of a semiconductor in equilibrium. Electronic states are described by delocalized Bloch wave functions. (Center) In the presence of an electric field, Bands are tilted and wave functions are interrupted at crossing points with the band edges. Wave functions are described now by Airy functions, characterised by a leaking tail into the forbidden region of the gap and damped oscillations toward the alowed energy range. (Right) This situation permits now optical absortion processes at photon energies below the nominal gap of the semiconductor by light-assited tunelling across a triangular barrier whose width is a function of the field intensity.

FKOs are routinely observed in PR spectra of bare semiconductors and structures, resulting from electric fields present at free surfaces and/or interfaces. A detailed review of FKOs in PR can be found in [37] and refs. therein. We show an example of FKOs and its standard analysis in Figure 12, which refers to a sample consisting of 20x InAs QD layer stack in a p-i-n structure of GaAs. Evidence of signatures attributed to QD-ground state and wetting layers are highlighted in Figure 12 (left), indicating also the energy of the GaAs barrier and the region of FKOs above its fundamental gap. Figure 12 (center) shows an enlarged view of the region of interest for FKO-analysis, together with an interpolation performed in order to improve the quality of the fit. The theoretical analysis of Aspnes for the medium-field regime described in a previous section leads to a relationship predicting the oscillatory behavior of PR spectra depending on the electric field intensity [7]:

$$\frac{\Delta R}{R} \sim \cos\left\{\frac{4\sqrt{2\mu}(E-E_g)^{3/2}}{3q\hbar F} + \pi\frac{d-1}{2}\right\},$$

where $\mu$ is the effective mass in the direction of the field $F$ and $d$ represents the dimensionality of the critical point from which the oscillations develop. Considering only the argument in the cosine function and after rearranging terms, we can plot the energy of the oscillation extrema $j$ (assigning $j=1,2,...$ to maxima and minima of the oscillations measured in the spectrum) as a function of a new variable $x_j$, defined as:

$$x_j = \left[\frac{3\pi\left(j-\frac{d}{2}\right)}{2}\right]^{2/3},$$

and a linear relationship follows between $E_j$ and $x_j$:

$$E_j = \left(\frac{q^2 F^2 \hbar^2}{8\mu}\right)^{1/3} x_j + E_g.$$

The linear relationship predicted between these two variables is reflected in the fit (dotted line) included in Figure 12 (right). From the linear fit, the slope is related to the intensity of the field (about 35 kV/cm for the 20x InAs QD layer stack contained in the intrinsic GaAs region of the p-i-n structure) and the ordinate at origin corresponds to the nominal bandgap of the semiconductor sustaining the field (1.42 eV, in excellent agreement with the fundamental gap of GaAs).

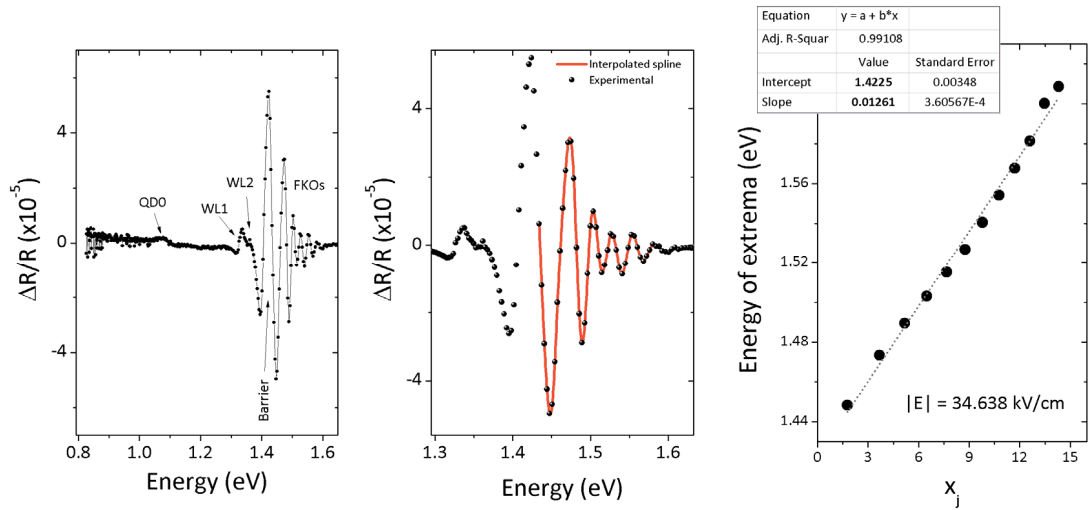

**Figure 12**: (Left) PR spectrum of a 20x InAs-QD layer stack embedded in the intrinsic region of a p-i-n GaAs host. QD and wetting layer (WL) signatures appear at energies below the bandgap of the GaAs barrier, above which Franz-Keldysh oscillations (FKOs) develop. (Center) Enlarged view of the energy range of FKOs (dots: experimental; solid line: spline interpolation). (Right) Linear dependence of extrem energy as a function of the corrected extram index $x_j$ (dots: experimental; dashed line: linear fit) and related fitting parameters.

Measuring the intensity of electric fields present in depletion regions of semiconductors is not straightforward. PR can provide access to the magnitude of such fields under conditions of medium-field regime, a significant advantage over alternative characterisation methods and particularly useful for the analysis of p-n junction-based devices such as solar cells.

### *4.6 Photoreflectance as diagnostic tool: correlating growth conditions and photovoltaic performance in multijunction solar cells.*

Multijunction and organic solar cells have shown the steepest increase in efficiency records over the last decade. While the former have already demonstrated efficiencies as high as 43.5% for a lattice-matched, triple-junction structure, the latter has just recently broken the 10% limit [38], still far from high efficiency figures. A critical issue in the optimization of triple-junction solar cells is the delicate definition of high quality interfaces between the varieties of materials that constitute the device. For instance, particularly critical is the change in growth conditions required to deposit the AlInP window layer on the GaInP top-cell, as to ensure optimal opto-electronic performance of both parts in the complete device.

Cánovas et al. have recently demonstrated that PR is capable to reveal the entire electronic structure of multijunction devices, including the determination of the critical points and absorption thresholds of each of the sub-cells completing the structure, as well as the electric field strengths sustained at the interfaces between them [39]. In this case we illustrate a different application of PR as a diagnostic tool. Figure 13 shows a set of PR measurements recorded around the absorption threshold of a GaInP top cell of three structures grown by MOVPE under different switching conditions for the growth of the AlInP/GaInP heterointerface. Essentially, the difference between the samples is the duration of the interruption in the Indium supply at the heterointerface. In sample 1 (black line in Fig. 13) the Indium supply is never interrupted and at the window/emitter interfaces Ga-supply is turned off and Al-supply is turned on; in sample 2 (red line in Fig. 13) the Indium supply is interrupted for 6 seconds and halfway in this interruption (i.e., after 3 s) Ga-supply is is turned off and Al-supply is turned on; finally, in sample 3 (green line in Fig. 13) the Indium supply is interrupted for 10 s and halfway in this interruption (i.e., after 5 s) Ga-supply is turned off and Al-supply is turned on. PR measurements were performed on these samples using the 325 nm laser line chopped at 777 Hz, 180 W QTH lamp and a Si-detector. Figure 13 (right) shows the corresponding external quantum efficiency (EQE) of the top cells of complete devices from the same batches. The effect of varying switching conditions on the properties of the GaInP top cell is evident in each figure. Furthermore, a direct correlation is found between the electronic quality of the device as measured from EQE and the intensity of the corresponding PR spectra. Such a direct correlation opens the possibility of using PR as a diagnostic tool for quality control, either post-growth or even in-situ during processing, if appropriate calibrations could be established between overall currents (from integrated QE-curves) and PR-intensities and/or broadening factors of the corresponding absorber layer.

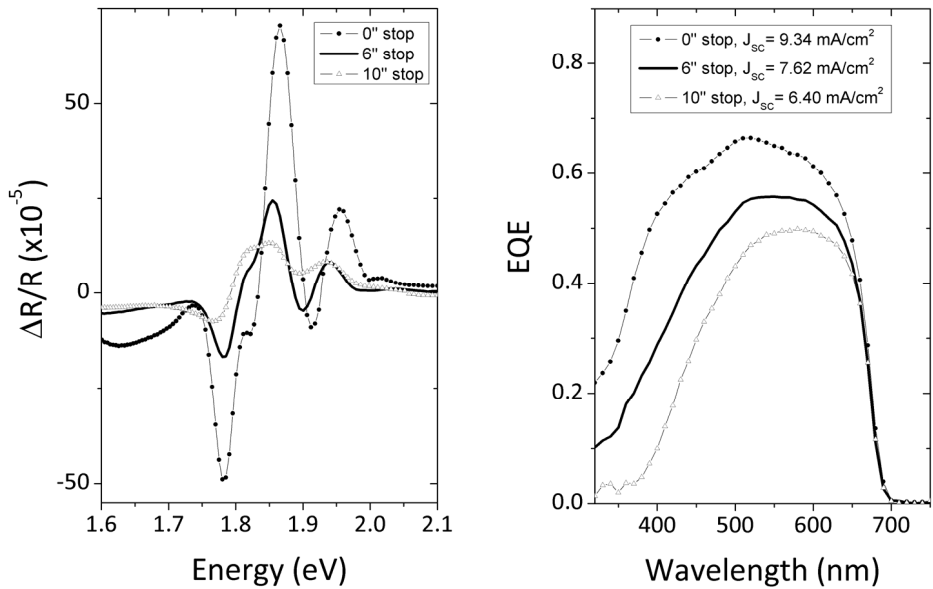

**Figure 13**: (Left) PR spectra around the absorption onset of GaInP used in top cells of multijunction photovoltaic devices grown by MOVPE under three different switching conditions for the growth of the window layer. (Right) Corresponding external quantum efficiencies of the top cells.

5. **Conclusions**

We have presented results of the characterization of different PV-structures by means of photoreflectance. The technique is able to provide useful information from different aspects of samples under study, including confinement effects, nature of the critical points associated with absorption thresholds, distinct electronic signatures of semiconductor alloys, and the magnitude of electric fields present in space-charge regions, among others. With such a wide range of accessible information, PR appears as a powerful tool for both fundamental research of new PV-materials and structures and for material diagnostics at early stages of industrial processing.

**Acknowledgements**

Financial support from the European Commission (project NGCPV - 283798), the Spanish Ministry of Economy & Competiveness (projects Nanogeffes – ENE2009-14481-C02-01, CERVINO – TEC2009-11143, and CELSO42 – TEC2011-28639-C02-01) and the Madrid Regional Government (project Numancia-II – S-2009/ENE1477) is gratefully acknowledged.

**References**

[1] M. Cardona, *Modulation spectroscopy*, Academic Press (1969).
[2] F.H. Pollak, Surf. Interface Anal. 31 (2001) 938.
[3] This experiment could be realized in practice by modulating the aperture of the exit slit of a monochromator in a wavelength-modulated reflectance measurement, an alternative modulation spectroscopy method.
[4] See, for example, L.P. Avakyants, P.Y. Bokov, A.V. Chervyakov, Tech. Phys. 50 (2005) 1316.
[5] J. Plaza, D. Ghita, J.L. Castaño, B.J. García, J. Appl. Phys. 102 (2007) 093507.
[6] $\varepsilon_{jj}$ refers the diagonalized components of the dielectric tensor along principal axes labeled jj.
[7] D.E. Aspnes in: *Handbook of Semiconductors*, vol. 2, edited by T.S. Moss, Elsevier (1994) p. 109.
[8] B.O. Seraphin, N. Bottka, Phys. Rev. 139 (1965) A560.
[9] O.J. Glembocki, SPIE Proc. Series, vol. 1286 (1990) 2.
[10] D.E. Aspnes, J.E. Rowe, Solid State Commun. 8 (1970) 1145.
[11] D.E. Aspnes, J.E. Rowe, Phys. Rev. B. 5 (1972) 4022.
[12] D.E. Aspnes, Surf. Sci. 37 (1973) 418.
[13] F.H. Pollak, O.J. Glembocki, Proc. SPIE 946 (1988).
[14] F.H. Pollak, in *Handbook of Semiconductors*, vol. 2, edited by T.S. Moss, Elsevier (1994) p. 527.
[15] A. Luque, A. Martí, Phys. Rev. Lett. 78 (1997) 5014.
[16] W. Shockley, H.J. Queisser, J. Appl. Phys. 32 (1961) 510.
[17] C.R. Stanley et al., in *Next Generation of Photovoltaics*, edited by A.B. Cristóbal, A. Martí, A. Luque, Springer Series in Optical Sciences 165 (2012) p. 251.
[18] E. Cánovas, A. Martí, N. López, E. Antolín, P.G. Linares, C.D. Farmer, C.R. Stanley, A. Luque, Thin Solid Films 516 (2008) 6943.
[19] O.J. Glembocki, B.V. Shanabrook, N. Bottka, W.T. Beard, J. Comas, Appl. Phys. Lett. 46 (1985) 970.
[20] J. Misiewicz, P. Sitarek, G.Sek, R. Kudrawiec, Mater. Sci. 21 (2003) 263.
[21] K.-H. Lee, K.W.J. Barnham, J.P. Connolly, B.C. Browne, R.J. Airey, J.S. Roberts, M. Führer, T.N.D. Tibbits, N.J. Ekins-Daukes, IEEE J. Photovolt. 2 (2012) 68.

[22] R. Oshima, T. Hashimoto, H. Shigekawa, Y. Okada, J. Appl. Phys. 100 (2006) 083110.
[23] W. Shan, W. Walukiewicz, J.W. Ager III, E.E. Haller, J.F. Geisz, D.J. Friedman, J.M. Olson, S.R. Kurtz, Phys. Rev. Lett. 82 (1999) 1221.
[24] W. Walukiewicz, W. Shan, K.M. Yu, J.W: Ager III, E.E. Haller, I. Miotlowski, M.J. Seong, H. Alawadhi, A.K. Ramdas, Phys. Rev. Lett. 85 (2000) 1552.
[25] W. Shan, W. Walukiewicz, K.M. Yu, J.W. Ager III, E.E. Haller, J.F. Geisz, D.J. Friedman, J.M. Olson, S.R. Kurz, K. Nauka, Phys. Rev. B 62 (2000) 4211.
[26] P.Y. Yu, M. Cardona, *Fundamentals of semiconductors*, Springer-Verlag (1996), p. 323.
[27] S. Turcotte, S. Larouche, J.-N. Beaudry, L. Martinu, R.A. Masut, P. Desjardins, R. Leonelli, Phys.Rev. B 80 (2009) 085203; N. Ashan, N. Miyashita, M.M. Islam, K.M. Yu, W. Walukiewicz, Y. Okada, Appl. Phys. Lett. 100 (2012) 172111.
[28] M. Geddo, T. Ciabattoni, G. Guizzetti, M. Galli, M. Patrini, A. Polimeni, R. Trotta, M. Capizzi, G. Bais, M. Piccin, S. Rubini, F. Martelli, A. Franciosi, Appl. Phys. Lett. 90 (2007) 091907.
[29] R. Kudrawiec, P. Sitarek, J. Misiewicz, S.R. Bank, H.B. Yuen, M.A. Wistey, J.S. Harris Jr. Appl. Phys. Lett. 86 (2005) 091115.
[30] R. Kudrawiec, M. Motyka, M. Gladysiewicz, P. Sitarek, J. Misiewicz, Appl. Surf. Sci. 253 (2006) 266.
[31] J. Shao, X.Lü, S. Guo, W. Lu, L. Chen, Y. Wei, J. Yang, L. He, Phys. Rev. B 80 (2009) 155125.
[32] D. Huang, D. Mui, H. Morkoç, J. Appl. Phys. 66 (1989) 358.
[33] S. Ghosh, B.M. Arora, J. Appl. Phys. 81 (1997) 6968.
[34] K.-S. Lee, J. Korean Phys. Soc. 49 (2006) 2045.
[35] W. Franz, Z. Naturforschg. 13 (1958) 484.
[36] L.V. Keldysh, J. Exptl. Theoret. Phys. (U.S.S.R.) 33 (1957) 994, translated in Soviet Phys. JETP 6 (1958) 763.
[37] H. Shen, M. Dutta, J. Appl. Phys. 78 (1995) 2151.
[38] http://www.nrel.gov/ncpv/images/efficiency_chart.jpg, accessed June, 15[th] 2012.
[39] E. Cánovas, D. Fuertes Marrón, A. Martí, A. Luque, A.W. Bett, F. Dimroth, S.P. Philipps, Appl. Phys. Lett. 97 (2010) 203504.